\documentclass[dvips]{article}
\usepackage{supertabular,lscape,epsfig}
\usepackage{amssymb}
\usepackage{amsmath}

\usepackage{pslatex}

\DeclareSymbolFont{ppa}{OT1}{ppl}{m}{it}
\DeclareMathSymbol{\vv}{\mathalpha}{ppa}{'166}

\begin{document}

\newcommand{\dd}{\,{\rm d}}
\newcommand{\ie}{{\it i.e.},\,}
\newcommand{\etal}{{\it et al.\ }}
\newcommand{\eg}{{\it e.g.},\,}
\newcommand{\cf}{{\it cf.\ }}
\newcommand{\vs}{{\it vs.\ }}
\newcommand{\zdot}{\makebox[0pt][l]{.}}
\newcommand{\up}[1]{\ifmmode^{\rm #1}\else$^{\rm #1}$\fi}
\newcommand{\dn}[1]{\ifmmode_{\rm #1}\else$_{\rm #1}$\fi}
\newcommand{\upd}{\up{d}}
\newcommand{\uph}{\up{h}}
\newcommand{\upm}{\up{m}}  
\newcommand{\ups}{\up{s}}
\newcommand{\arcd}{\ifmmode^{\circ}\else$^{\circ}$\fi}
\newcommand{\arcm}{\ifmmode{'}\else$'$\fi}
\newcommand{\arcs}{\ifmmode{''}\else$''$\fi}
\newcommand{\MS}{{\rm M}\ifmmode_{\odot}\else$_{\odot}$\fi}
\newcommand{\RS}{{\rm R}\ifmmode_{\odot}\else$_{\odot}$\fi}
\newcommand{\LS}{{\rm L}\ifmmode_{\odot}\else$_{\odot}$\fi}

\newcommand{\Abstract}[2]{{\footnotesize\begin{center}ABSTRACT\end{center}
\vspace{1mm}\par#1\par   
\noindent
{~}{\it #2}}}

\newcommand{\TabCap}[2]{\begin{center}\parbox[t]{#1}{\begin{center}
  \small {\spaceskip 2pt plus 1pt minus 1pt T a b l e}
  \refstepcounter{table}\thetable \\[2mm]
  \footnotesize #2 \end{center}}\end{center}}

\newcommand{\TableSep}[2]{\begin{table}[p]\vspace{#1}
\TabCap{#2}\end{table}}

\newcommand{\FigCap}[1]{\footnotesize\par\noindent Fig.\  %
  \refstepcounter{figure}\thefigure. #1\par}

\newcommand{\TableFont}{\footnotesize}
\newcommand{\TableFontIt}{\ttit}
\newcommand{\SetTableFont}[1]{\renewcommand{\TableFont}{#1}}

\newcommand{\MakeTable}[4]{\begin{table}[htb]\TabCap{#2}{#3}
  \begin{center} \TableFont \begin{tabular}{#1} #4
  \end{tabular}\end{center}\end{table}}

\newcommand{\MakeTableSep}[4]{\begin{table}[p]\TabCap{#2}{#3}
  \begin{center} \TableFont \begin{tabular}{#1} #4
  \end{tabular}\end{center}\end{table}}

\newenvironment{references}%
{
\footnotesize \frenchspacing
\renewcommand{\thesection}{}
\renewcommand{\in}{{\rm in }}
\renewcommand{\AA}{Astron.\ Astrophys.}
\newcommand{\AAS}{Astron.~Astrophys.~Suppl.~Ser.}
\newcommand{\ApJ}{Astrophys.\ J.}
\newcommand{\ApJS}{Astrophys.\ J.~Suppl.~Ser.}
\newcommand{\ApJL}{Astrophys.\ J.~Letters}
\newcommand{\AJ}{Astron.\ J.}
\newcommand{\IBVS}{IBVS}
\newcommand{\PASP}{P.A.S.P.}
\newcommand{\Acta}{Acta Astron.}
\newcommand{\MNRAS}{MNRAS}
\renewcommand{\and}{{\rm and }}
\section{{\rm REFERENCES}}
\sloppy \hyphenpenalty10000
\begin{list}{}{\leftmargin1cm\listparindent-1cm
\itemindent\listparindent\parsep0pt\itemsep0pt}}%
{\end{list}\vspace{2mm}}
 
\def\TYLDA{~}
\newlength{\DW}
\settowidth{\DW}{0}
\newcommand{\dw}{\hspace{\DW}}

\newcommand{\refitem}[5]{\item[]{#1} #2%
\def\REFARG{#3}\ifx\REFARG\TYLDA\else, {\it#3}\fi
\def\REFARG{#4}\ifx\REFARG\TYLDA\else, {\bf#4}\fi
\def\REFARG{#5}\ifx\REFARG\TYLDA\else, {#5}\fi.}

\newcommand{\Section}[1]{\section{#1}}
\newcommand{\Subsection}[1]{\subsection{#1}}
\newcommand{\Acknow}[1]{\par\vspace{5mm}{\bf Acknowledgements.} #1}
\pagestyle{myheadings}

\newfont{\bb}{ptmbi8t at 12pt}
\newcommand{\xrule}{\rule{0pt}{2.5ex}}  
\newcommand{\xxrule}{\rule[-1.8ex]{0pt}{4.5ex}}  
\def\thefootnote{\fnsymbol{footnote}}
\newcommand{\uprule}{\rule{0pt}{2.5ex}}
\newcommand{\douprule}{\rule[-2ex]{0pt}{4.5ex}}
\newcommand{\dorule}{\rule[-2ex]{0pt}{2ex}}

\begin{center}
{\Large\bf
Evolution of Low Mass Contact Binaries}
\vskip1.7cm
{\bf 
K.~~ S~t~\c e~p~i~e~\'n}$^!$~~ and {\bf K.~~G~a~z~e~a~s}$^{2,3}$
\vskip7mm
{$^1$Warsaw University Observatory, Al.~Ujazdowskie~4, 00-478~Warsaw, Poland\\
email: kst@astrouw.edu.pl\\
$^2$Department of Astrophysics, Astronomy and Mechanics, University of
Athens, GR-157 84, Zografos, Athens, Greece\\
$^3$European Space Agency, ESTEC, Mechatronics and Optics
Division, Keplerlaan~1, 2200AG, Noordwijk, The Netherlands\\
email: kgaze@physics.auth.gr, kosmas.gazeas@esa.int}
\end{center}

\Abstract{We present a study on low-mass contact binaries (LMCB) with 
orbital periods shorter than 0.3 days and total mass lower than about
1.4~\MS. We show that such systems have a long pre-contact phase, which
lasts for 8--9~Gyrs, while the contact phase takes only about 0.8~Gyr, which
is rather a short fraction of the total life. With low mass transfer rate
during contact, moderate mass ratios prevail in LMCBs since they do not
have enough time to reach extreme mass ratios often observed in higher mass
binaries. During the whole evolution both components of LMCBs remain within
the MS band.

The evolution of cool contact binaries towards merging is controlled by the
interplay between the evolutionary expansion of the less massive component
resulting in the mass transfer to the more massive component and the
angular momentum loss from the system by the magnetized wind. In LMCB the
angular momentum loss prevails. As a result, the orbital period
systematically decreases until the binary overflows the outer critical
Roche surface and the components merge into a single fast rotating star of
a solar type surrounded by a remnant disk carrying excess angular
momentum. The disk can be a place of planet formation with the age
substantially lower than the age of a host star. The calculated theoretical
tracks show good agreement with the physical properties of LMCB from the
available observations. Estimates of the frequency of occurrence of LMCB
and the merger formation rate indicate that about 40 LMCBs and about 100
low mass merger products is expected to exist within 100 pc from the Sun.}
{Planets and satellites: formation -- Stars: contact -- Stars:
eclipsing -- Stars: binary -- Stars: evolution -- Stars: W~UMa}
\vspace*{-7pt}
\Section{Introduction -- Properties of Contact Binaries}
\vspace*{-5pt}
W~UMa-type binaries are composed of two cool stars in contact with each
other, surrounded by a common convective envelope lying between the inner
and outer critical Roche surfaces (Mochnacki 1981). In spite of different
component masses they possess almost identical surface brightness. The more
massive component of a typical W~UMa-type system is a main sequence (MS)
star lying not far from the zero age MS (ZAMS) whereas the lower mass
component is oversized, sometimes by a factor of several, compared to its
expected ZAMS radius (St\c epie\'n 2006). They occur in our solar neighborhood
with an apparent frequency of one binary among about 500 single stars in
the Galactic disk (Rucinski 2006), but their frequency in stellar clusters
depends on the cluster age. They have never been observed in open clusters
younger than 0.7~Gyr but are ubiquitously found in globular or old open
clusters. Their spatial frequency can be as high as 1/45 among blue
stragglers in the globular clusters (Rucinski 2000).

W~UMa-type systems are considered to be low temperature contact binaries
with components of F, G, K spectral type. Hot, massive contact binaries of
O and B spectral types are also known but they are surrounded by the common
radiative envelope and they seem to have a different origin.

Many thousands of W~UMa-type systems are presently known with the majority
of them discovered photometrically by the automated sky surveys like OGLE
(Szymański, Kubiak and Udalski 2001) or ASAS (Pojmański 2002). However,
only about 150 systems have been studied systematically up to date,
combining good quality photometric and spectroscopic observations of which
a dozen or so has periods shorter than 0.3~d. These data have been
published in our older studies. In the current study, they have been
updated for a few systems with the new, recently obtained observational
data, while a few more systems have been added as a result of the ongoing
observational program of W~UMa-type stars, in order to increase the sample
and cover the entire range of their properties. Following strict selection
criteria, only contact binaries with accurate combined solutions based on
high quality photometric and radial velocity data for both components have
been included in our study. They are listed in Table~1.  Solutions based
solely on the photometric observations, where the mass ratio is estimated,
are not considered to give reliable results. However, we list such binaries
in Table~2 as supplementary data.  The system GSC~1387:0475 is the only
exception in Table~2, as its mass ratio has been determined
spectroscopically by Rucinski and Pribulla (2008). They also give the
approximate photometric solution stressing, however, that the reliable
values of the binary parameters are very difficult to obtain due to a very
low orbital inclination of the system. Indeed, the solution given by Yang,
Wei and Li (2010) is so much different from the one obtained by Rucinski
and Pribulla (2008) that we decided to put the star in Table~2.
\renewcommand{\arraystretch}{0.95}
\renewcommand{\TableFont}{\scriptsize}
\MakeTable{l@{\hspace{3pt}}c@{\hspace{4pt}}c@{\hspace{7pt}}c@{\hspace{7pt}}c@{\hspace{7pt}}c@{\hspace{7pt}}c@{\hspace{7pt}}c@{\hspace{7pt}}c@{\hspace{7pt}}c@{\hspace{7pt}}c@{\hspace{0pt}}c@{\hspace{0pt}}c}{12.5cm}{The list of LMCBs used in the current study, with
accurately determined physical parameters (combined photometric
and spectroscopic solution)}
{\hline
\noalign{\vskip3pt}
Name & $P_{\rm orb}$ & $\log P$ & $q_{\rm sp}$ & $M_{\rm pr}$ & $M_{\rm sec}$ & $R_{\rm pr}$ & $R_{\rm sec}$ & $L_{\rm pr}$ & $ L_{\rm sec}$ & a     &  $H_{\rm orb}{\cdot}10^{-51}$ &  Ref \\
     & [days]        &          &              & [\MS]        & [\MS]         & [\RS]        & [\RS]         & [\LS]        & [\LS]          & [\RS] &  [g$\cdot$cm$^2$/s]                 & \\
\noalign{\vskip3pt}
\hline
\noalign{\vskip3pt}
CC~Com          &   0.2207  & $-0.6562$ &   0.529   &   0.720   &   0.379   &   0.708   &   0.522   &   0.151   &   0.080   &   1.585   &   1.985   &    1      \\
V523~Cas        &   0.2337  & $-0.6313$ &   0.533   &   0.740   &   0.381   &   0.728   &   0.536   &   0.139   &   0.104   &   1.658   &   2.096   &    1      \\
RW~Com          &   0.2373  & $-0.6247$ &   0.471   &   0.800   &   0.380   &   0.770   &   0.540   &   0.260   &   0.149   &   1.720   &   2.202   &    2      \\
VZ~Psc          &   0.2613  & $-0.5829$ &   0.800   &   0.810   &   0.650   &   0.780   &   0.700   &   0.220   &   0.124   &   1.970   &   3.678   &    3      \\
44~Boo          &   0.2678  & $-0.5722$ &   0.488   &   0.940   &   0.470   &   0.840   &   0.590   &   0.573   &   0.348   &   2.000   &   3.123   &    4      \\
V345~Gem        &   0.2748  & $-0.5610$ &   0.142   &   1.144   &   0.163   &   1.103   &   0.427   &   1.781   &   0.313   &   1.944   &   1.371   &    5      \\
BD~07$^{o}$3142 &   0.2752  & $-0.5604$ &   0.662   &   0.740   &   0.490   &   0.810   &   0.670   &   0.269   &   0.229   &   2.045   &   2.729   &    2      \\
BX~Peg          &   0.2804  & $-0.5522$ &   0.372   &   1.030   &   0.383   &   0.957   &   0.610   &   0.639   &   0.308   &   2.014   &   2.853   &    6      \\
XY~Leo          &   0.2841  & $-0.5465$ &   0.717   &   0.813   &   0.593   &   0.833   &   0.714   &   0.448   &   0.220   &   2.036   &   3.498   &    1      \\
RW~Dor          &   0.2854  & $-0.5445$ &   0.703   &   0.711   &   0.499   &   0.832   &   0.622   &   0.274   &   0.181   &   1.944   &   2.719   &    7      \\
BW~Dra          &   0.2922  & $-0.5343$ &   0.280   &   0.920   &   0.260   &   0.980   &   0.550   &   1.085   &   0.386   &   1.964   &   1.853   &    8      \\
OU~Ser          &   0.2968  & $-0.5275$ &   0.173   &   1.109   &   0.192   &   1.148   &   0.507   &   1.460   &   0.341   &   2.041   &   1.612   &    9      \\
TZ~Boo          &   0.2972  & $-0.5270$ &   0.207   &   0.990   &   0.210   &   1.080   &   0.560   &   1.240   &   0.342   &   1.982   &   1.593   &    10     \\
\hline
\noalign{\vskip3pt}
\multicolumn{13}{p{12.5cm}}{
Index `pr' and `sec' stands for the currently observed primary (more
massive) and secondary (less massive) component respectively, while the
mass ratio is defined as $q=M_{\rm sec}/M_{\rm pr}$.\newline
1.~Zola \etal (2010), 2.~Djurasević \etal (2011), 3.~Hrivnak, Guinan and Lu
(1995) 4.~Maceroni \etal (1981), 5.~Gazeas (in preparation), 6.~Lee \etal
(2009), 7.~Marino \etal (2007), 8.~Kaluzny and Rucinski (1986), 9.~Zola
\etal (2005), 10.~Christopoulou, Papageorgiou and Chrysopoulos (2011)}}
\renewcommand{\TableFont}{\footnotesize}
\MakeTableSep{l@{\hspace{23pt}}c@{\hspace{23pt}}c@{\hspace{23pt}}cc}{12.5cm}{The supplemented list of LMCBs used in the current study,
for which the mass ratio was determined photometrically}
{\hline
\noalign{\vskip3pt}
Name       &$P_{\rm orb}$ &$\log P$ &  $q_{\rm ph}$  &  Ref  \\
           &[days]         &         &                 &     \\
\noalign{\vskip3pt}
\hline
\noalign{\vskip3pt}
GSC~1387:0475       &   0.2178  & $-0.6619$ &   0.474      & 1   \\
YZ~Phe              &   0.2347  & $-0.6295$ &   0.402      & 2   \\
BI~Vul              &   0.2518  & $-0.5989$ &   0.692      & 3   \\
V803~Aql            &   0.2634  & $-0.5794$ &   1.000      & 4   \\
FS~CrA              &   0.2636  & $-0.5791$ &   0.758      & 3   \\
BP~Vel              &   0.2650  & $-0.5768$ &   0.532      & 5   \\
EK~Com              &   0.2667  & $-0.5740$ &   0.300      & 6   \\
V802~Aql            &   0.2677  & $-0.5724$ &   0.160      & 7   \\
IV~Dra              &   0.2680  & $-0.5719$ &   0.500      & 8   \\
DD~Com              &   0.2692  & $-0.5699$ &   0.271      & 9   \\
FG~Sct              &   0.2706  & $-0.5677$ &   0.786      & 10  \\
BM~UMa              &   0.2712  & $-0.5667$ &   0.540      & 11  \\
V743~Sgr            &   0.2766  & $-0.5581$ &   0.319      & 12  \\
VW~Cep              &   0.2783  & $-0.4445$ &   0.400      & 13,14\\
V1005~Her           &   0.2790  & $-0.5544$ &   0.290      & 15  \\
AD~Cnc              &   0.2827  & $-0.5487$ &   0.770      & 16  \\
V524~Mon            &   0.2836  & $-0.5473$ &   0.442      & 17  \\
ER~Cep              &   0.2857  & $-0.5441$ &   0.549      & 18  \\
V700~Cyg            &   0.2906  & $-0.5367$ &   0.637      & 19  \\
V676~Cen            &   0.2924  & $-0.5340$ &   0.720      & 20  \\
V902~Sgr            &   0.2939  & $-0.5318$ &   0.120      & 21  \\
EH~Hya              &   0.2969  & $-0.5274$ &   0.314      & 10  \\
ASAS~050837+0512.3  &   0.2660  & $-0.5751$ &   0.709   &   22  \\
ASAS~155906-6317.8  &   0.2670  & $-0.5735$ &   0.500   &   22  \\
ASAS~052452-2809.2  &   0.2760  & $-0.5591$ &   0.250   &   22  \\
ASAS~095048-6723.3  &   0.2770  & $-0.5575$ &   0.500   &   22  \\
ASAS~033959+0314.5  &   0.2830  & $-0.5482$ &   0.709   &   22  \\
ASAS~212915+1604.9  &   0.2830  & $-0.5482$ &   0.353   &   22  \\
ASAS~195350-5003.5  &   0.2870  & $-0.5421$ &   0.709   &   22  \\
ASAS~061531+1935.4  &   0.2880  & $-0.5406$ &   0.353   &   22  \\
ASAS~085710+1856.8  &   0.2910  & $-0.5361$ &   0.353   &   22  \\
ASAS~143751-3850.8  &   0.2920  & $-0.5346$ &   0.500   &   22  \\
ASAS~120036-3915.6  &   0.2930  & $-0.5331$ &   0.250   &   22  \\
ASAS~144243-7418.7  &   0.2950  & $-0.5302$ &   0.250   &   22  \\
ASAS~050852+0249.3  &   0.2960  & $-0.5287$ &   0.353   &   22  \\
ASAS~155227-5500.6  &   0.2980  & $-0.5258$ &   0.709   &   22  \\
ASAS~174655+0249.9  &   0.3000  & $-0.5229$ &   0.353   &   22  \\
\noalign{\vskip3pt}
\hline
\noalign{\vskip3pt}
\multicolumn{5}{p{11cm}}{
1.~Rucinski and Pribulla (2008), 2.~Samec and Terrell (1995), 3.~Bradstreet
(1985), 4.~Samec, Su and Dewitt (1993), 5.~Lapasset, Gomez and Farinas
(1996), 6.~Samec, Carrigan and Padgen (1995), 7.~Samec, Martin and Faulkner
(2004), 8.~Robb (1992), 9.~Zhu \etal (2010), 10.~Maceroni and van't Veer
(1996), 11.~Samec, Gray and Garrigan (1995), 12.~Samec, Carrigan and Wei
Lool (1998), 13.~Khajavi, Edalati and Jassur (2002), 14.~Binnendijk (1966),
15.~Robb, Greimel and Oulette (1997), 16.~Qian \etal (2007), 17.~Samec and
Loflin (2004), 18.~Maceroni, Milano and Russo (1984), 19.~Niarchos,
Hoffmann and Duerbeck (1997), 20.~Gray, Woissol and Samec (1996), 21.~Samec
and Corbin (2001), 22.~Pilecki and St\c epie\'n (2012).}}

All W~UMa-type show very similar observational and physical
properties. Apparently, they have the same origin and evolution, as shown
by Gazeas and Niarchos (2006) and Gazeas and St\c epie\'n (2008). According to
Bilir \etal (2005) the stars have a kinematic age of 5--12~Gyr. This is in
agreement with theoretical models, explaining the formation of contact
systems by the systematic angular momentum loss (AML) in initially detached
binaries with orbital periods of a couple of days, due to the magnetized
stellar winds and tidal coupling (Vilhu 1981, 1982, Rahunen 1981, 1982,
1983, St\c epie\'n 1995, 2006). Although evolution of binaries towards the
contact configuration is quite clear, the evolution during the contact
phase and beyond is still controversial, since AML, and mass and energy
path exchange between the components lead to different evolution than the
one we currently know for single stars. The thermal relaxation oscillation
(TRO) model, described by Lucy (1976), Flannery (1976) and Webbink (1977),
and later developed by a number of authors (Yakut and Eggleton 2005 and
references therein), assumes that each component of the binary is out of
thermal equilibrium and its size oscillates around the inner Roche
lobe. The low mass component is oversized due to the convective energy
transfer from the massive component. The binary spends a part of its life
in contact and the rest as a semi-detached binary, slowly evolving towards
an extreme mass ratio system. An alternative evolutionary model of cool
contact binaries has been developed by St\c epie\'n (2004, 2006, 2009) and will
be presented in more detail in Section~3. Briefly, the model assumes that
mass transfer occurs with the mass ratio reversal, similarly as in
Algol-type binaries, following the Roche lobe overflow (RLOF) by the
massive component. The contact configuration is formed immediately after
that or after some additional AML. Each component is in thermal equilibrium
and the large size of the currently less massive component results from its
advanced evolutionary stage (its core is hydrogen depleted). The energy
flows from the presently massive component in a form of the large scale,
steady circulation of high entropy matter bound to the equatorial region
and encircling the low mass component.

Crucial for the further evolution in contact is the ratio of the AML time
scale to the evolutionary time scale of the initially more massive
component (St\c epie\'n 2011). Its value determines the evolutionary stage of
the massive component when it reaches RLOF and transfers rapidly mass to
its companion. We consider three different cases: the massive component
reaches RLOF when it is still on MS but not yet approaching TAMS, when it
is close to, or slightly beyond TAMS, and when it is on the subgiant
branch, approaching the red giant branch. Note that no sharp limit occurs
between the first and second case, as well as between the second and third
case. The transition is continuous so the assignment to the particular case
results from the evolutionary behavior following RLOF. In analogy with the
terminology introduced by Kippenhahn and Weigert (1967) for the upper MS
binaries we will call Case A, AB and B the three above described
situations.

Case~A rapid mass exchange results in a binary composed of two MS stars
with the presently less massive component (donor) being more advanced
evolutionary than the accretor. That happens for low mass binaries with
short initial periods. After the rapid mass transfer in Case~AB the donor
is completely, or nearly completely depleted of hydrogen in its center but
it has not yet built a significant helium core. It leaves MS and enters the
subgiant branch. This is associated with a relatively faster expansion than
during the MS phase, which results in a higher mass transfer rate. This
prevents the orbit from a rapid shrinking caused by AML and gives the donor
enough time to transfer almost all its mass to the accretor. The mass ratio
reaches the extreme value leading to the Darwin instability and merging of
both components. In Case~B RLOF occurs when the star is already approaching
the red giant branch and possesses a small helium core of a few tenths of
the solar mass. This happens in binaries with long orbital periods. The
rapidly expanding star fills its critical Roche lobe when the orbit is
still wide and the rapid mass exchange results in formation of an
Algol-type binary with the donor being a red giant or subgiant.

In this study we concentrate on contact binaries formed as a result of
Case~A mass transfer in binaries with the initial period between 1.5 and
2.5~d, (St\c epie\'n 2011), and the initial mass of the massive component low
enough that its MS life time is longer than the AML time scale. Contact
binaries formed following RLOF in Case~A will be called low mass contact
binaries (LMCB). The slow expansion of the primary results in a moderate
mass transfer and thus does not lead to widening of the orbit. In the long
run AML prevails, leading to the overflow of the outer critical Roche lobe
and the coalescence of both components when the mass ratio is still
moderate.

\Section{The Low Mass Contact Binaries}
LMCBs can be distinguished observationally by their very short orbital
periods ($P_{\rm orb}\leq0.3$~d), moderate mass ratios ($0.2\leq q\leq0.8$
with a strong concentration around 0.5) low component masses (individually
between 0.2--1~\MS\ and totally between 1.0--1.4~\MS) and radii equal to
the MS objects. There may be a few exceptions with parameters outside of
these limits, nevertheless the great majority of LMCBs have parameters
within the above ranges. Their components have late spectral types (G to K)
and consequently they show low surface temperatures (4000--5000~K). Such
stars are expected to be magnetically very active with cool star spots
covering a significant -- but strongly variable -- fraction of their
surface. As a result, their light curves vary on a time scale of years,
showing often asymmetry known as O'Connell Effect. In some rare cases their
variation is noticed within a few months or even weeks (Gazeas, Niarchos
and Gradoula 2006). In almost all LMCBs the massive component shows lower
apparent surface brightness than its companion. This is explained by a
large coverage of its photosphere with cool, dark spots reducing
significantly its apparent luminosity (Eaton, Wu and Rucinski 1980, St\c epie\'n
1980, Hendry, Mochnacki and Collier Cameron 1992).  Photometrically this
effect illustrates the so-called W-phenomenon (Binnendijk 1970) according
to which the photometric minimum of light variation occurs when the less
massive component is eclipsed, not the more massive one, as normally
expected (A-phenomenon). The vast majority of LMCB shows the W-phenomenon,
which often confuses the observers on which component is the more massive
and which is the less massive one.

The undisturbed photometric temperature of the more massive component is
very likely higher than the temperature of the low mass component, as the
recent modeling of high quality observations shows (Zola \etal
2010). Because the spottiness varies in time, some stars may migrate from
W-type to A-type and {\it vice versa}. Apparently, this migration does not occur
among LMCB or occurs only exceptionally.

In agreement with their mass and their late G and K spectral type, LMCB are
intrinsically faint. This is the reason why the number of such systems with
well determined parameters is very limited. Rucinski (2007) argues,
however, that the period distribution of the volume limited sample of
W~UMa-type systems has a maximum around 0.27 days, \ie within the range of
LMCB. So, in reality, they must be quite numerous in space. Also, according
to the results by Bilir \etal (2005) they belong to the oldest population
of contact binaries, with the kinematic age of 9~Gyr. Recently, Pilecki
(2009) obtained photometric solutions for about 2900 eclipsing binaries
with good light curves from the sky survey ASAS (Pojmański 2002). To
automate the process, he solved each light curve for several pre-defined
values of mass ratio and selected the solution best representing the
observed curve. Because of the relatively coarse pre-defined grid, the
obtained mass ratios of the analyzed binaries are not distributed
continuously but, instead, are grouped at several specified values. In
spite of the low accuracy of the mass ratio determinations, his results for
contact binaries with periods shorter than 0.3~d show that they all fit
well into our criteria for LMCB: their mass ratios lie between 0.25 and 0.8
(see Table~2) and they all show the W-phenomenon.

\Section{Theoretical Background and the New Model}
Several theoretical models were proposed to explain the characteristics of
contact binaries (CBs), with only a few able to explain the most
characteristic observational data. There are several assumptions in these
models, which were made mainly due to the lack of observational evidence or
a very limited sample of analyzed CBs. Independent sets of evolutionary
models of single stars, produced by many authors, show significant
discrepancies, particularly in radius \vs time, time scales to reach a
given evolutionary stage \eg to build a helium core of a given mass
etc. Such uncertainties must also exist among programs modeling evolution
of close binaries towards the contact phase.

Nevertheless, all models show that both components of a close binary,
evolving later into a CB, should lie inside their Roche lobes for some time
at early evolutionary phases. When the orbital AM becomes low enough, one
or both components will eventually overflow their inner Roche lobes,
resulting in a contact configuration with a common envelope around the
whole system.

Cool CBs are among the binary systems with the highest level of
chromospheric-coronal activity. Therefore, they lose AM and mass by the
magnetized wind, just as it is observed in single, highly active stars. In
addition, the existing models of CB assume that the low mass component
(donor) transfers mass to its companion (accretor) on the evolutionary time
scale. The relative rates of all these processes determine the course of
binary evolution till merging of both components. In particular, the mass
transfer, resulting from the evolutionary expansion of the donor, acts in
the direction of the orbit widening. In the absence of AML, the orbital
period increases at a rate determined by the condition that the donor
overfills its Roche lobe by a narrow margin. The resulting mass transfer
rate is rather low, as observed in classical Algols with periods longer
than a week, where the AML rate by the magnetized wind is low.  However,
CBs with periods shorter than 1 day have high AML rates, which changes the
situation radically. The AML acts in the direction of shrinking the orbit
and both Roche lobes. When the contracting Roche lobe moves deeper beneath
the donor surface, the mass transfer rate increases, widening the orbit
again. The equilibrium rates of both processes depend on many details of
which very important is the evolutionary advancement of the donor. In LMCBs
its expansion results from evolution across MS, which is rather slow, hence
AML dominates bringing the binary to coalescence before the mass ratio
becomes very low. In binaries which are past mass exchange in Case AB or B,
donors expand much faster forcing a high mass transfer rate. This prevents
the binary from overflowing the outer critical surface in the early phase
of the mass transfer process, so the merging of both components will occur
only when the extreme mass ratio $q\leq0.1$ is reached.

Based on the observational data and the fact that the TRO model encounters
several difficulties (Webbink 2003, St\c epie\'n 2011, and references therein) a
new, alternative model has been suggested by St\c epie\'n (2006, 2009; see also
Gazeas and St\c epie\'n 2008). The main features of this model are:
\vspace*{-11pt}
\begin{enumerate}
\parskip=0pt \itemsep=1mm \setlength{\itemsep}{0.4mm}\setlength{\parindent}{-1em} \setlength{\itemindent}{0em}
\item Cool CBs evolve from detached binaries with
initial periods close to 2~days.
\item Both components are magnetically active, at the
highest, so called saturation, level.
\item Strong magnetized winds blow from both components leading to mass and
AM loss. With the full spin-orbit synchronization, the orbital AM is
ultimately reduced.
\item Evolutionary expansion of the massive component, together
with orbit shrinkage, results in RLOF
followed by the rapid mass transfer to the low mass component.
\item The rapid mass exchange proceeds until the mass ratio reversal, similarly
as in Algol-type binaries.
\item Depending on the detailed values of the orbital parameters, either a
CB emerges directly from the rapid mass exchange phase, or a near-contact
binary is formed, which reaches contact after some additional AML.
\item Further evolution of the binary in contact proceeds under the influence
of AML and slow mass transfer from the present, evolutionary
advanced, low mass component (donor) to the present massive component
(accretor).
\item At the end, both components merge together forming a single,
rapidly rotating star.
\end{enumerate}

Perhaps the most important feature of the present model, differing it from
TRO model, is listed under item~(5). TRO model assumes that the matter
flowing from the more massive, hence hotter, component covers the low mass
component completely. The blanket of high entropy gas blocks its core
energy flux, which instead of being radiated into space, heats up the
convection zone until the specific entropy equals that of the covering
matter. The increased entropy makes the convection zone expand until the
star fills its Roche lobe. It was shown, however, that the flow of matter
from the hotter component is a dynamical process (Tassoul 1992, St\c epie\'n
2009). The resulting stream is bound to the equatorial region by the
Coriolis force and the low mass component can radiate freely its core
energy from the polar regions not covered by the hot matter. Its global
parameters, including the radius, change very little, compared to a single
star of the same mass. Apart from a temporary loss of thermal equilibrium,
the rapid mass transfer following RLOF by the massive component of a close
binary will result in a reversal of the mass ratio.

Starting from the initial conditions the following set of equations is
solved for each time step
\vspace*{-11pt}
\begin{eqnarray}
\frac{\dd H_{\rm orb}}{\dd t}&=&-4.9\times10^{41}(R_1^2M_1+R_2^2M_2)/P\,,\\
\dot M_{1,2}&=&-10^{-11}R_{1,2}^2\,,\\
P&=&0.1159\cdot a^{3/2}M^{-1/2}\,,
\end{eqnarray}
\begin{eqnarray}
H_{\rm orb}&=&1.24\times10^{52}M^{5/3}P^{1/3}q(1+q)^{-2}\,,\\
\frac{r_1}{a}&=&\frac{0.49\cdot q^{2/3}}{0.6\cdot q^{2/3}+\ln{(1+q^{1/3})}}\,,\\
\frac{r_2}{a}&=&\frac{0.49\cdot q^{-2/3}}{0.6\cdot q^{-2/3}+\ln{(1+q^{-1/3})}}\,.
\end{eqnarray}

Here $t$ is time in years, $P$, $H_{\rm orb}$ and $a$ are the orbital
period (in days), orbital AM (in cgs units) and semi-axis (in solar units),
$M_{1,2}$ and $R_{1,2}$ are masses and radii of both components also in
solar units, and $M=M_1+M_2$. Just for reference, the solar spin AM is
$H_\odot=1.63\times 10^{48}$ in cgs units. From now on, the subscript '1'
will denote the initially more massive component and the subscript '2' the
initially less massive component so that the initial mass ratio $q\equiv
M_1/M_2>1$ but after the rapid mass exchange $q<1$. To allow for the effect
of the activity supersaturation $P\equiv0.4$~d is adopted in Eq.~(1) for
periods shorter than 0.4~days. The relations $R_1(t)$ and $R_2(t)$, taken
from the evolutionary models of single stars computed by Girardi \etal
(2000) and Sienkiewicz (see St\c epie\'n 2006), supplement the model.

Eqs.~(5--6) give approximate sizes of the Roche lobes of both components
$r_1$ and $r_2$ (Eggleton 1983). For contact configurations these sizes are
assumed to be identical with stellar radii.

We assume in the current model that cool CB systems evolve from initially
detached configurations with stellar parameters corresponding to ZAMS and
orbital periods close to 2~days. Evolution is split in three phases: the
detached phase~I, until the RLOF by star '1', the rapid mass exchange
phase~II and the slow mass transfer phase~III, which lasts until the
presumed merging of the components. To simplify computations, the constant
mass transfer rate in phase~III is assumed. It is so adjusted that the
evolutionary radius of the donor (star '1') is close to the size of its
Roche lobe during the whole phase~III.

Nothing spectacular happens during phase~I -- the system simply loses mass
and AM {\it via} the winds while the components slowly expand due to their
evolution across the MS. Phase~I is described for several binaries in
detail by St\c epie\'n (2011). Phase~II begins when the period is well below
1~day. This phase is very short compared to two other phases. Some mass
(and AM) may be lost during it but allowing for them adds two more free
parameters to the model, which we want to avoid, so the present
calculations are made assuming the conservative mass exchange. Then the
system enters phase~III when the stars reach contact and evolve as a
contact binary.

\Section{The Results of Calculations and Comparison with Observations}
The observations put constraints on the initial parameter values of the
binaries which evolve later into LMCB. The presently observed total mass
range of LMCB restricts the values of the initial total mass to the range
1.1--1.6~\MS, because the expected total mass loss due to the winds is
around 0.1--0.2~\MS. The observed values of the donor radii (star '1')
suggest that the stars are close to TAMS and hydrogen in their cores is
substantially depleted. This means that they must have been close to TAMS
already at RLOF because the following evolutionary phases last much shorter
than phase~I (see Table~3). This requires that $M_{1,{\rm init}}$ can not
be lower than 0.9~\MS\ for the metal abundance characteristic of old disk
and lower than 0.8~\MS\ if the star is metal-poor. Keeping in mind, on the
other hand, that the initial total mass should be less than 1.6~\MS\ we put
a condition $M_{1,{\rm init}}\le1.1$~\MS\ to avoid extreme initial mass
ratios for which the assumption of the conservative mass transfer in
phase~II is most likely not fulfilled.

A set of evolutionary models of cool close binaries with different initial
conditions was calculated by one of us (St\c epie\'n 2011). Time evolution of 27
new models was followed, with initial orbital periods of 1.5, 2.0 and
2.5~days and initial components masses 1.3+1.1, 1.3+0.9, 1.3+0.7, 1.1+0.9,
1.1+0.7, 1.1+0.5, 0.9+0.7, 0.9+0.5 and 0.9+0.3 (in solar units). Table~3
gives the details of the models evolving into LMCBs. After the rapid mass
exchange (phase~II) the constant mass transfer rate was adopted for
phase~III. Its value was adjusted to keep the radius of the donor within a
few percent of the size of its Roche lobe at the beginning and the end of
phase~III. The donor radius was interpolated using the set of evolutionary
models of single stars with different masses. Taking into account the
decrease of the donor mass and its evolutionary effects during phase~III,
the value of the radius at each time step was obtained from the
interpolation in time and mass. In most cases the donor radius deviated
from the size of its Roche lobe by no more than 10\% over the whole
phase~III. In a few cases, however, keeping a constant mass transfer rate
over the whole phase~III caused the donor radius to deviate from the size
of its Roche lobe by more than 10\%. In these cases, phase~III was divided
into two parts with two different mass transfer rates. End of phase~III was
signaled by a condition $q<0.1$ or when the binary overflowed the outer
critical surface.

As it turned out, the most massive binaries with $P_{\rm init}= 2.5$~d
evolve into short-period Algols with donors possessing a small helium
core. This is so because phase~I takes more time than the MS life time of
$M_1$ in these binaries (St\c epie\'n 2011). RLOF takes place when the donor has
already left MS on its way to the red giant branch, which results in a high
mass transfer rate. After phase~II it continues a fast expansion and AML
rate is insufficient to prevent the binary from orbit widening, so it
evolves as classical Algols with constantly increasing period. Binaries
with $P_{\rm init}=2$~d evolve into CBs with lower mass transfer rates
resulting from the fact that their donors have been just at, or very close
to TAMS at RLOF and do not expand so fast as in previous models. Binaries
with $P_{\rm init}=1.5$~d reach RLOF when $M_1$ is still at a distance from
TAMS, so after the rapid mass exchange both components reside on MS.  Those
are the most compact binaries (particularly when both component masses are
low) and with a low mass transfer rate in phase~III they merge quickly.

We compare a subset of the above described models, fulfilling the condition
$M\le1.6$~\MS, with the observations of LMCB. Below, we discuss briefly all
such models and we show that only some of them satisfactorily reproduce the
observations. These models are then discussed in more details. In the
following we denote models with $M_{1,{\rm init}}+M_{2,{\rm init}}(P_{\rm
init})$. Apart from the models presented in St\c epie\'n (2011) two
additional models characteristic of LMCB have been calculated. These are:
0.9+0.4(2.5) and 0.8+0.3(2.5). The latter binary was evolved assuming a
decreased metal content $Z=0.001$.

\begin{itemize}
\parskip=0pt \itemsep=1mm \setlength{\itemsep}{1.3mm}\setlength{\parindent}{-1em} \setlength{\itemindent}{0em}
\item Model 0.9+0.3(1.5). The model has the lowest initial AM of all
  considered. Due to a short period, AML rate is relatively high (see
  Eq.~1). RLOF occurs after 3.8~Gyr when $P=0.27$~d and star '1' is
  hardly evolved from ZAMS. Very soon after the beginning of the rapid mass
  exchange the binary overflows the outer critical Roche lobe due to the
  further fast period shortening. A very high mass and AM loss
  follows, resulting in coalescence of both components. The model never
  reaches phase III and no stable contact configuration is formed.

\item Model 0.9+0.3(2.0). The model has an evolutionary history very
  similar to the previous one. RLOF occurs after 4.7~Gyr when star '1' has
  not yet reached 50\% of its MS life. Rapid mass transfer tightens the
  orbit during phase II so much that the binary overflows the outer
  critical surface. Again, the model never forms a stable contact
  configuration.

\item Model 0.9+0.3(2.5). The model has a significantly higher initial AM
  than both previous models and, at the same time, lower AML rate. RLOF
  occurs after 8.6~Gyr when $P=0.35$~d and the star '1' reaches about
  70\% of its MS life (it would be more if its metal content were lower than
  solar). The thermal equilibrium radii of both components are always
  smaller than the outer critical surface during phase II so the binary can
  survive it and enter phase III. Its details are given in 
  Table~3.

\item Model 0.9+0.4(2.5). This is the most massive model with the initial
  period of 2.5~d which evolves into LMCB with properties similar to observed.
  Listed in Table~3.

\item Model 0.9+0.5(1.5). Due to the higher initial AM, compared
  to the four previous models, it takes 5.5~Gyr until RLOF, in spite of
  the very short initial period. The period at RLOF is equal to 0.33~d. 
  Both equilibrium radii are comfortably within the outer critical
  surface during the rapid mass exchange (the minimum period at phase~II is
  equal to 0.25~d). Listed in Table~3.

\item Model 0.9+0.5(2.0). RLOF occurs after 9~Gyr when $P=0.39$~d and
  star '1' reaches about 70\% of its MS life. The orbit is wide enough
  during phase~II and the binary enters phase~III after the rapid mass
  exchange. Listed in Table~3.

\item Model 0.9+0.5(2.5). High initial AM, together with a low AML rate
  results in a very long phase~I. RLOF occurs only after 13~Gyr. This is
  much longer than the age of the Galactic disk so the model is rejected
  from the further discussion.

\item Model 0.9+0.7(1.5). RLOF occurs after 6~Gyr when $P=0.35$~d
and star '1' is almost halfway across MS. In binaries with the
mass ratio not far from unity, phase II lasts significantly longer
than in binaries with other mass ratios as in this
model where it takes 0.35 Gyr, compared to about 0.1 Gyr
in other models. At the end of phase II, when the equilibrium
configuration of star '1' recedes into its Roche lobe, star '2'
lies well inside its Roche lobe, so we obtain a near contact
binary of the Algol type. However, AML dominates in phase III and
the period decreases fast. Soon a contact configuration is formed
which lasts for a short time, ending with the overflow of the
outer critical surface. Listed in Table~3.

\item Model 0.9+0.7(2.0). RLOF occurs after 10 Gyr when $P=0.41$~d and
  star '1' is close to TAMS. Similarly as in the previous case, star '2' is
  within its Roche lobe after phase~II, so we obtain again a near contact
  binary of the Algol type. The system evolves into a contact
  configuration after about 0.5~Gyr. The period shortens until the overflow
  of the outer critical surface occurs. Listed in Table~3.

\item Model 0.9+0.7(2.5). The initial period is too long for RLOF to occur
  within the age of Universe so the model was not discussed any more.

\item Model 1.1+0.5(1.5). RLOF occurs after 3.4~Gyr when $P=0.39$~d and
  star '1' is about halfway across MS. After additional 0.1~Gyr a contact
  configuration is formed which exists for almost 1~Gyr. Listed in 
  Table~3.

\item Model 1.1+0.5(2.0). RLOF occurs after 5.5~Gyr when $P=0.45$~d and
  star '1' approaches TAMS. Contact is established soon after wards and
  it lasts for almost 1~Gyr. Listed in Table~3.

\item Model 1.1+0.5(2.5). RLOF occurs after 6.8~Gyr when $P=0.91$~d and
  star '1' is already a subgiant with a significant helium core. The binary
  evolves into Algol with systematically increasing period. The model was
  excluded from further discussion.

\item Model 0.8+0.3(2.5). This is the model with the lowest total mass
  considered by us. In contrast to the previous models we assumed here the
  decreased metal content $Z=0.001$, so that star '1' evolves faster than
  the solar composition model of the same mass. RLOF occurs after 11~Gyr
  when $P=0.47$~d and star '1' approaches TAMS. Following the rapid mass
  transfer a near contact binary is formed, which evolves into a contact
  configuration after additional 0.7~Gyr. It stays in contact for 0.6~Gyr
  until coalescence occurs. Listed in Table~3.
\end{itemize}

\renewcommand{\arraystretch}{1}
\MakeTableSep{@{}lccclc@{}}{12.5cm}{Results of model calculations}
{\hline 
\noalign{\vskip3pt}
       Evolutionary    & Age    & $M_1+M_2$  & $q$     &  $P_{\rm orb}$  & $H_{\rm orb}$        \\
          stage        & [Gyr]  & [\MS]      &         &  [days]         & ($\times 10^{51}$)    \\
\noalign{\vskip3pt}
\hline
\noalign{\vskip3pt}
Initial (ZAMS) & 0 & 0.9+0.3 & 3.0 & 2.50 & 4.275\\
Start RLOF & 8.64 & 0.835+0.292 & 2.86 & 0.350 & 2.051\\
Start contact & 8.74 & 0.355+0.772 & 0.46 & 0.230 & 2.001\\
Coalescence & 9.42 & 0.251+0.867 & 0.29 & 0.219 & 1.571\\
&&&&&\\
Initial (ZAMS) & 0 & 0.9+0.4 & 2.25 & 2.50 & 5.550\\
Start RLOF & 11.40 & 0.816+0.382 & 2.14 & 0.426 & 2.740\\
Start contact & 12.20 & 0.390+0.797 & 0.49 & 0.287 & 2.401\\
Coalescence & 13.00 & 0.210+0.961 & 0.22 & 0.286 & 1.562\\
&&&&&\\
Initial (ZAMS) & 0 & 0.9+0.5 & 1.80 & 1.50 & 5.709\\
Start RLOF & 5.46 & 0.858+0.487 & 1.76 & 0.327 & 3.230\\
Start contact & 5.66 & 0.557+0.786 & 0.71 & 0.245 & 3.077\\
Coalescence & 6.24 & 0.440+0.895 & 0.49 & 0.207 & 2.623\\
&&&&&\\
Initial (ZAMS) & 0 & 0.9+0.5 & 1.80 & 2.00 & 6.284\\
Start RLOF & 8.89 & 0.832+0.478 & 1.74 & 0.385 & 3.278\\
Start contact & 9.39 & 0.480+0.823 & 0.58 & 0.267 & 2.888\\
Coalescence & 10.50 & 0.256+1.032 & 0.25 & 0.234 & 1.858\\
&&&&&\\
Initial (ZAMS) & 0 & 0.9+0.7 & 1.29 & 1.50 & 7.645\\
Start RLOF & 5.97 & 0.854+0.672 & 1.27 & 0.345 & 4.336\\
Start contact & 6.44 & 0.620+0.899 & 0.69 & 0.259 & 3.832\\
Coalescence & 6.71 & 0.565+0.950 & 0.59 & 0.228 & 3.540\\
&&&&&\\
Initial (ZAMS) & 0 & 0.9+0.7 & 1.29 & 2.00 & 8.414\\
Start RLOF & 10.2 & 0.825+0.653 & 1.26 & 0.408 & 4.348\\
Start contact & 10.9 & 0.524+0.942 & 0.56 & 0.288 & 3.558\\
Coalescence & 11.9 & 0.276+1.177 & 0.23 & 0.244 & 2.221\\
&&&&&\\
Initial (ZAMS) & 0 & 1.1+0.5 & 2.20 & 1.50 & 6.674\\
Start RLOF & 3.38 & 1.066+0.492 & 2.17 & 0.386 & 4.084\\
Start contact & 3.48 & 0.585+0.971 & 0.60 & 0.276 & 3.957\\
Coalescence & 4.46 & 0.287+1.255 & 0.23 & 0.289 & 2.559\\
&&&&&\\
Initial (ZAMS) & 0 & 1.1+0.5 & 2.20 & 2.00 & 7.346\\
Start RLOF & 5.45 & 1.045+0.487 & 2.15 & 0.452 & 4.201\\
Start contact & 5.84 & 0.497+1.029 & 0.48 & 0.292 & 3.651\\
Coalescence & 6.63 & 0.312+1.203 & 0.26 & 0.253 & 2.560\\
&&&&&\\
Initial (ZAMS) & 0 & 0.8+0.3 & 2.67 & 2.50 & 3.912\\
Start RLOF & 11.2 & 0.734+0.290 & 2.53 & 0.473 & 2.040\\
Start contact & 11.97 & 0.327+0.683 & 0.48 & 0.236 & 1.707\\
Coalescence & 12.55 & 0.263+0.735 & 0.36 & 0.201 & 1.407\\
\noalign{\vskip3pt}
\hline}

\begin{figure}[htb]
\vglue-3mm
\centerline{\includegraphics[width=11cm]{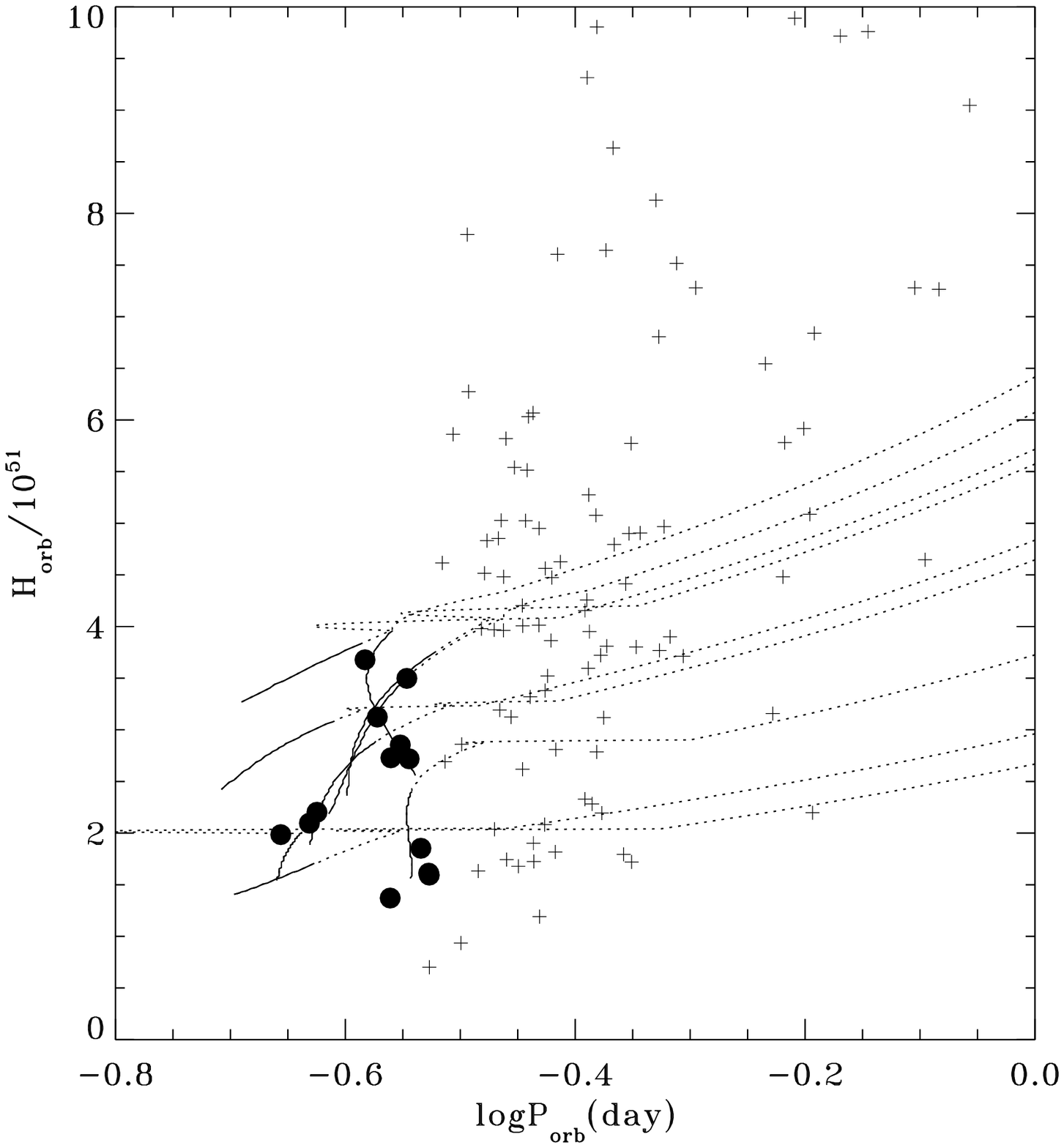}}
\vskip-17pt
\FigCap{Comparison of the observed orbital AM of
LMCBs from Table~1 -- filled circles, and observations of other W~UMa-type
systems listed in Gazeas and St\c epie\'n (2008) -- 'plus' signs, with the model
calculations from Table~3. Solid lines correspond to contact phase and
broken lines correspond to earlier phases of the following models (from top
to bottom): 0.9+0.7(1.5), 0.9+0.7(2.0), 1.1+0.5(1.5), 1.1+0.5(2.0),
0.9+0.5(1.5), 0.9+0.5(2.0), 0.9+0.4(2.5), 0.9+0.3(2.5), 0.8+0.3(2.5).}
\end{figure}
The above examples show that LMCBs are formed from binaries with the
initial values of parameters lying within quite narrow intervals. Binaries
with the low initial total mass -- less or equal to 1.3~\MS, and the initial
orbital periods between 1.5 and 2~days do not survive phase~II. RLOF occurs
when both components are still close to ZAMS so the binary is very compact
and the rapid mass exchange results in the immediate merger due to the
overflow of the outer critical surface followed by the very fast mass and
AM loss from the system. Low initial mass binaries evolve into LMCB only if
the initial orbital period is equal to 2.5~days. The binaries with longer
periods (of 3~days or more) have so low AML rates in phase~I that RLOF
occurs after the time longer than the Hubble time. Binaries with high total
initial mass of 1.4--1.6~\MS\ must have short initial orbital periods of
1.5--2~days for RLOF to occur at the right moment, \ie when star '1' is
close to TAMS. For initial orbital periods equal to 2.5~days or more, and
$M_{1,\rm init}=0.9$~\MS, the duration of phase~I in the high mass
binaries is again longer than the Hubble time. If, instead, $M_{1,\rm
init}=1.1$~\MS, RLOF occurs when $M_1$ is already on its way to the red
giant branch, so the rapid mass exchange results in the formation of an
Algol-type binary configuration.

\hglue-7pt It should be stressed again that the present model of the cool close binary
evolu\-tion does not contain any free parameters. Relaxing this restriction,
\eg by rejec\-ting the assumption of the conservative rapid mass exchange,
would certainly broa\-den the acceptable range of initial parameters leading
to the formation of LMCB.

\begin{figure}[htb]
\centerline{\includegraphics[width=11cm]{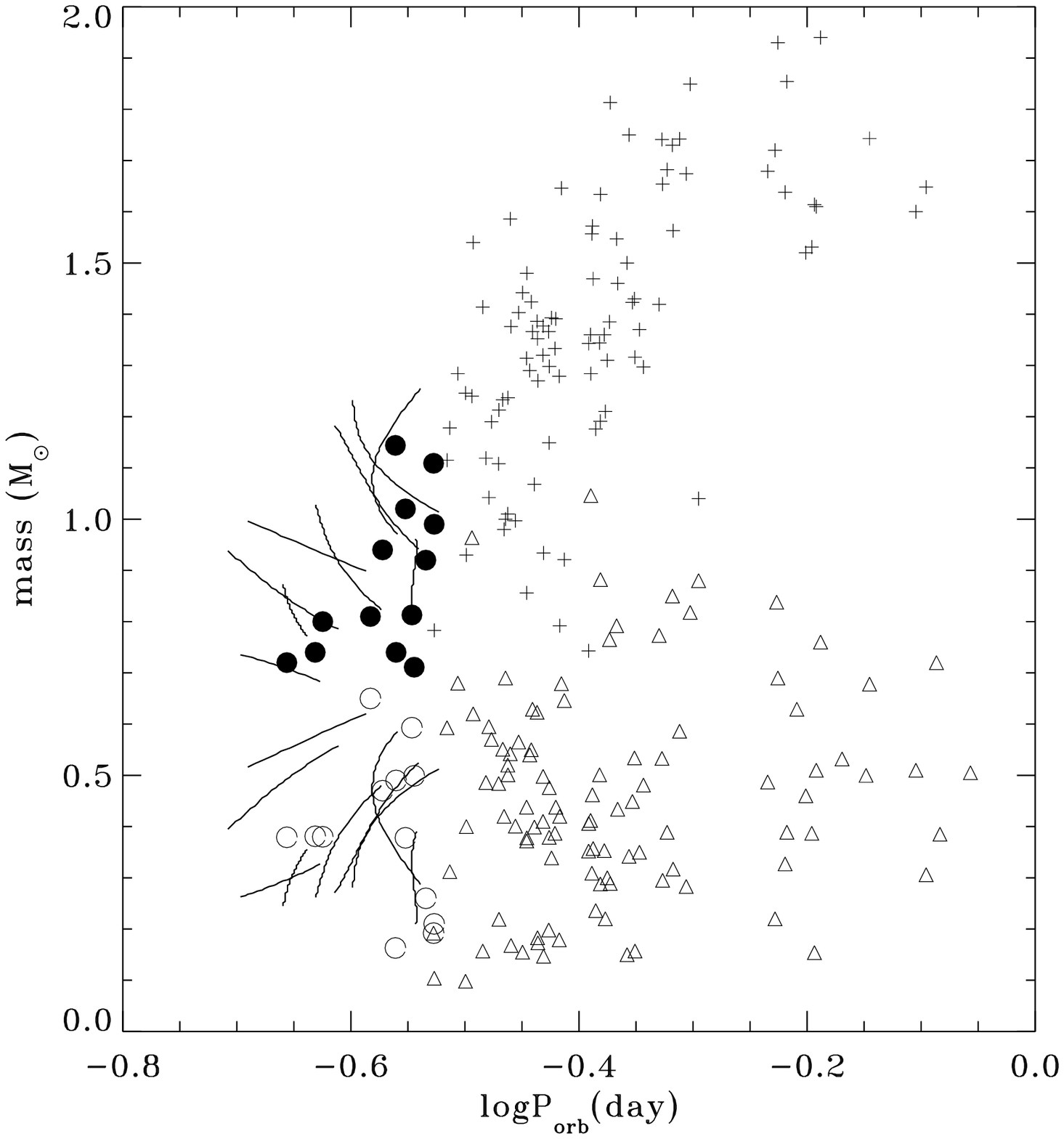}}
\vskip-17pt
\FigCap{The observed values of the component masses
of LMCBs from Table~1 (open and filled circles) and of other W~UMa-type
systems listed in Gazeas and St\c epie\'n (2008) ('plus' signs and triangles)
are compared with the model contact binaries listed in Table~3.}
\end{figure}
\begin{figure}[t]
\vglue-17pt
\centerline{\includegraphics[width=11cm]{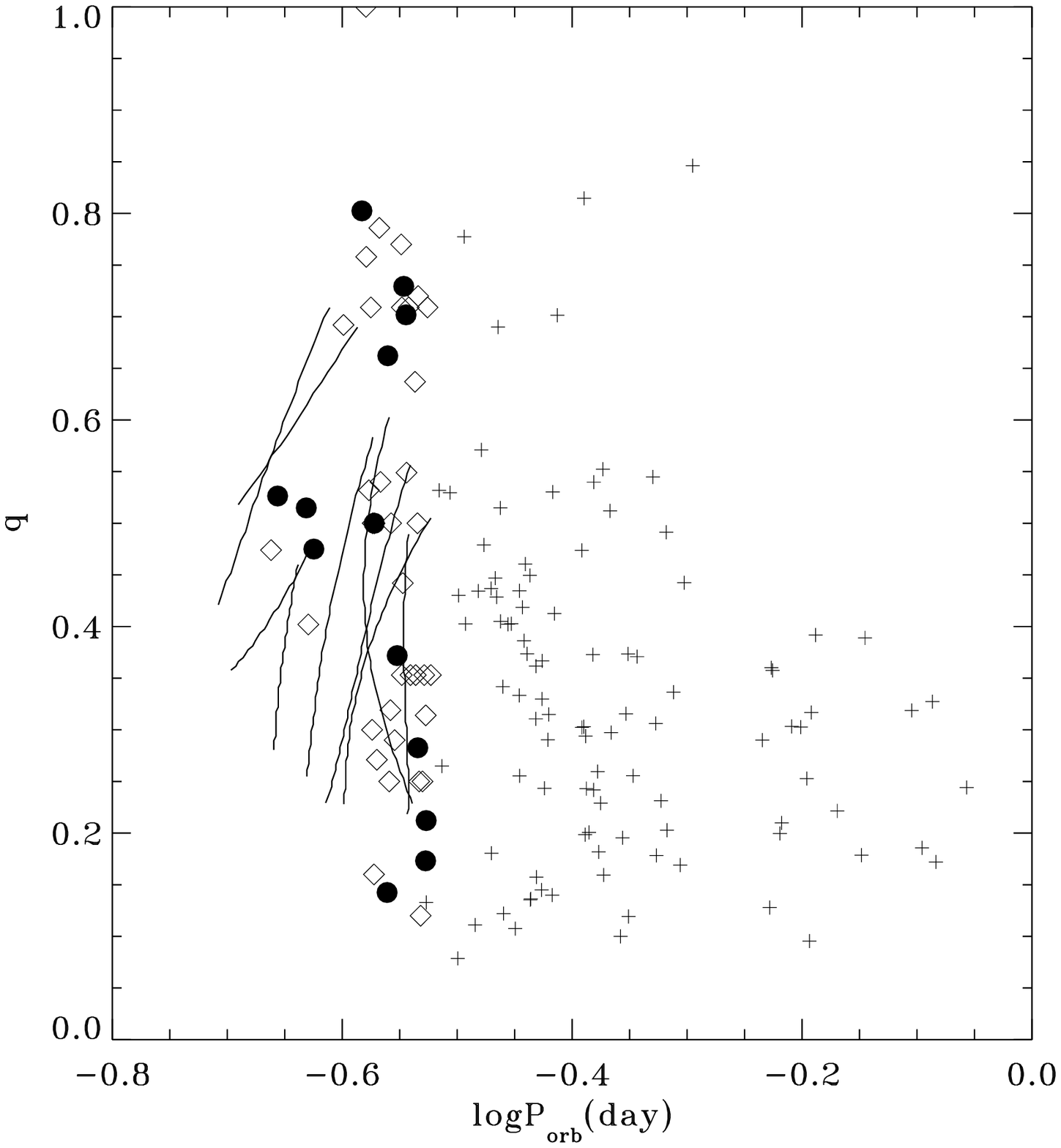}}
\vskip-17pt
\FigCap{The observed (spectroscopically determined)
values of the mass ratios for the LMCBs listed in Table~1 (filled circles),
supplemented by the photometrically determined mass ratios of the LMCBs
listed in Table~2 (diamonds) and the other W~UMa-type systems listed in
Gazeas and St\c epie\'n (2008) ('plus' signs) are compared with the evolutionary
models (the contact phase) listed in Table~3. Note that some of the diamond
signs are grouped at specific values, following the coarse pre-defined
search grid (see text).}
\end{figure}
\begin{figure}[htb]
\vglue-17pt
\centerline{\includegraphics[width=11cm]{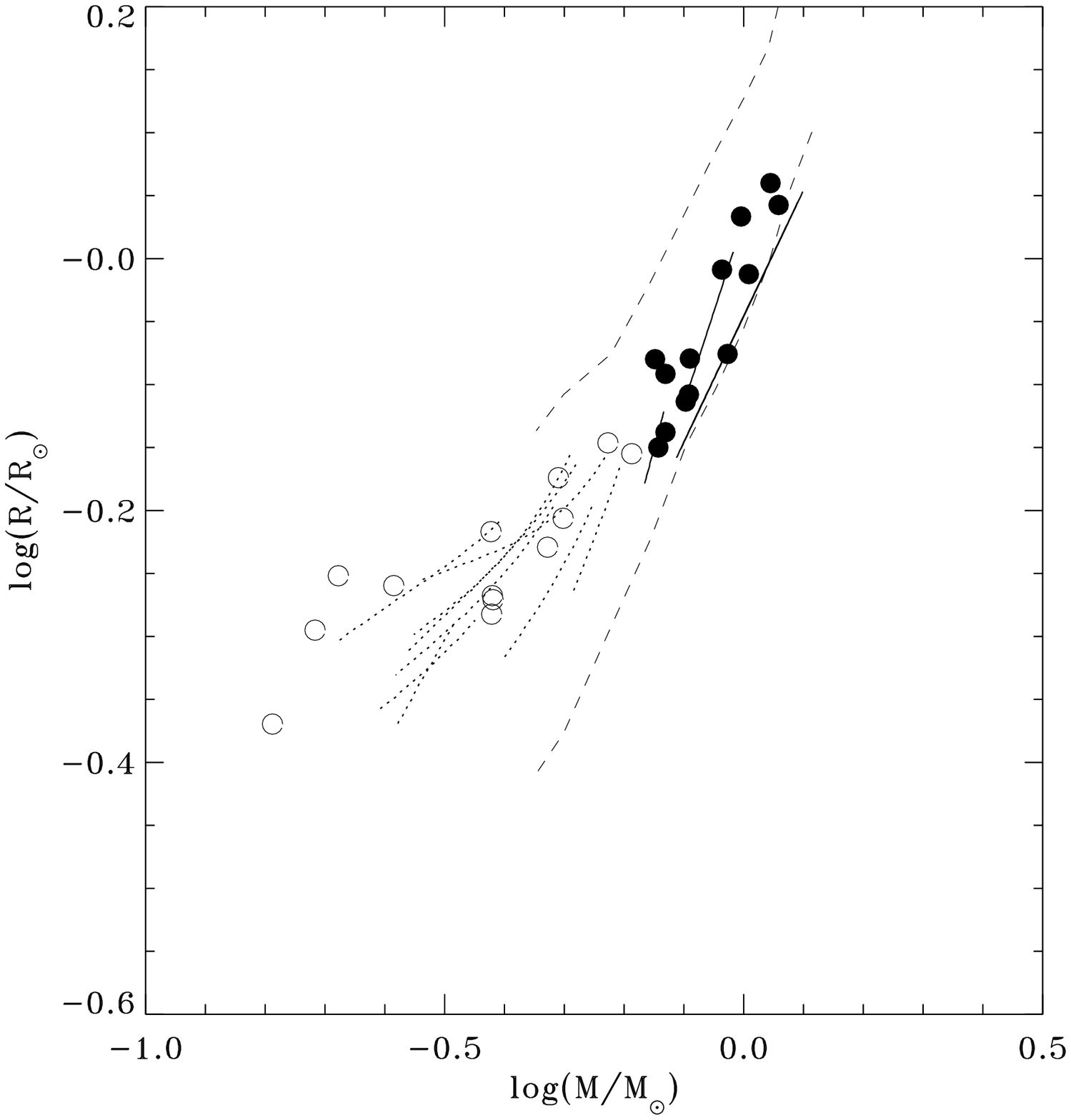}}
\vskip-17pt
\FigCap{The mass--radius diagram for the components
of the W~UMa-type systems from Table~1 (open and filled
circles). Evolutionary tracks of the models from Table~3 are overplotted as
solid and dotted lines. Note that several solid lines lying along ZAMS
overlap with one another. Broken lines show limits of the MS for the solar
composition models computed by Girardi \etal (2000).}
\end{figure}
Figs.~1--4 compare the models with observations. In Fig.~1 the observed
period-orbital AM diagram of LMCBs from Table~1 is shown (full circles),
together with AM of longer orbital period CBs listed in Gazeas and St\c epie\'n
(2008) ('plus' signs). Overplotted are lines showing the evolution of the
orbital AM of the models from Table~3 (dotted lines -- phases before
contact, solid lines -- contact configurations). The orbital period of
binaries with high initial (\ie before the reversal) mass ratios ($q
\approx2.5{-}3$) goes through the local minimum during phase~II. This is
shown in Fig.~1 as a loop or hook extending towards short orbital period
values.  The most prominent one reaches $\log P=-0.8$ and corresponds to
the model 0.9+0.3(2.5). Its extend suggests that the assumption of the
conservative rapid mass transfer may already be violated in this case. As
we see, the predicted values of the orbital AM in the contact phase
correspond closely to the observed values.

Fig.~2 shows the observed values of the component masses \vs orbital
period: open and filled circles correspond to more massive and less massive
components of LMCBs respectively, while 'plus' signs and triangles to the
other CBs of our sample. Lines describe the models in the contact
phase. Again, as we see, the agreement is excellent.

Fig.~3 gives the observed mass ratio as a function of period. Here, again,
the filled circles correspond to LMCBs from Table~1. To strengthen our
observation that LMCB avoid extreme mass ratios, we also plotted in Fig.~3
several binaries with periods shorter than 0.3~d and mass ratios determined
photometrically. These values are considered not to be very reliable, as
they lack spectroscopic confirmation or they result from low-quality
observational data. These data are given in Table~2 and are plotted as
diamonds. The concentration of $q$ -- values around 0.5 for the
shortest-period W~UMa-type stars has already been noted by Rucinski (2010)
in his Fig.~2. Note that none of the computed models extends beyond the
orbital period of 0.2~d, because all the considered binaries overflow the
outer critical Roche lobe before they reach this value. This prediction
agrees very well with the observed short-period limit of W~UMa-types stars
(Rucinski 2007). As in the previous figures, the observed values are well
reproduced by the models, shown here as solid lines.

Fig.~4 shows the mass--radius diagram for the high and low mass components
of LMCBs. They are marked with open and filled circles, respectively.
Overplotted are evolutionary tracks of models from Table~3. The broken
lines denote limits of MS taken from the paper by Girardi \etal
(2000). Note that the mass of the massive components increases during the
evolution in contact whereas the mass of the low mass components
decreases. Very good agreement of the observations with the predicted
values support the conclusion that LMCB are formed as a result of Case~A
mass exchange and they finish the evolution in contact when their
components are still on MS or, at most, just beyond TAMS.

\section{Discussion}
\subsection{Total Age and Duration of the Contact Phase}
The total age of the models from Table~3 varies from about 4.5~Gyr to
13~Gyr. It depends primarily on the initial mass of star '1'. To produce
LMCB, star '1' should be sufficiently advanced evolutionary at RLOF, \ie
its age should reach a considerable fraction of its total MS life time.
The latter value is equal to 6.4~Gyr for a 1.1~\MS\ star and about 15~Gyr
for an 0.9~\MS\ star (both for the solar composition). In effect, LMCBs
originated from binaries with $M_{1,\rm init}=1.1$~\MS\ cannot be older
than about 6~Gyr whereas those originating from binaries with $M_{1,\rm
init}=0.9$~\MS\ can be as old as even the Galactic disk. On the other hand,
the range of the initial binary parameters leading to LMCB narrows for
massive $M_{1,\rm init}$ stars. Consequently, only two such models are
listed in Table~3. These are: 1.1+0.5(1.5) and 1.1+0.5(2.0). Binaries with
$M_{1,\rm init}=1.1$~\MS\ and $M_{2,\rm init}>0.5$~\MS, or with $M_{2,\rm
init}=0.5$~\MS\ and original periods longer than 2.0 d do not evolve into
LMCB (note that binaries with $M_{1,\rm init}=1.1$~\MS\ and $M_{2,\rm
init}<0.5$~\MS\ were not considered here to avoid the non conservative
phase~II). For $M_{1,\rm init}=0.9$~\MS\ the range of the allowed parameters
leading to the formation of LMCBs is broader: $M_{2,\rm init}$ can vary
between 0.3 and 0.7~\MS, and the initial period between 1.5 and 2.5~d,
except that not all combinations of these parameters are permitted. In
particular, to obtain LMCB, the shorter initial period is needed for the
more massive $M_{2,\rm init}$. The average age of the models with $M_{1,\rm
init}=0.9$~\MS\ is 9.6~Gyr with a dispersion of about 50\%. The model
0.8+0.3(2.5) with the decreased metallicity behaves similarly as somewhat
more massive models with the solar metallicity.

The duration of the contact phase of the models from Table~3 varies from
about 0.3~Gyr to 1.1~Gyr with the average value of 0.6~Gyr. This value
depends on the adopted mass transfer rate in phase~III. As was mentioned
earlier, the mass transfer rate was adjusted so as to make sure that star
'1' is close to its Roche lobe at the beginning and the end of phase~III,
and then its value was kept constant over the whole phase~III (with a few
exceptions when two different values were used for the initial and final
part of phase~III). Obviously, the mass transfer rate in real binaries is
instantaneously adjusted to keep star '1' slightly exceeding its Roche lobe
all the time. The rates used in our model computations can be considered as
some sort of means over phase~III. They were equal to $1{-}3\times
10^{-10}$~\MS/yr, \ie about 2--3 times less than in other contact binaries
(Gazeas and St\c epie\'n 2008). The precise duration time of the contact phase
is very sensitive to some of the assumptions adopted in the present
investigation, as will be discussed in Section~5.2.

The contact phase takes from 4\% to 22\% of the total age. This percentage
is less than 10\% for LMCB originating from binaries with $M_{1,\rm
init}=0.9$~\MS\ and the highest values are for LMCB originating from
binaries with $M_{1,\rm init}= 1.1$~\MS\ \ie those with a relatively short
total age. In some of the considered models the contact phase is preceded
by a short semidetached phase when star '1' fills its Roche lobe and
transfers mass to star '2' on the evolutionary time scale (past phase~II)
but star '2' does not yet fill its Roche lobe. An additional AML is needed
to form a contact configuration. It is achieved in a fraction of
1~Gyr. This near contact phase can be shortened or completely avoided if a
necessary amount of AM is lost during the non conservative phase~II. If,
however, the rapid mass exchange is conservative, we predict the existence
of a population of semidetached binaries with cool components filling their
Roche lobe and periods close to but shorter than 0.3~d. Indeed, Pilecki
(2009) lists several such binaries detected within the framework of the
ASAS program.

Before comparing our final models with observations one should take into
account the expected distribution of the initial parameters. Let us assume
a flat distribution in $q$ and $P_{\rm orb}$ over the considered interval
of the initial parameters, and the Salpeter initial mass function for
$M_{1,\rm init}$. A rough estimate of the relative frequency of LMCB
originating from binaries with $M_{1,\rm init}=1.1$~\MS\ and from binaries
with $M_{1,\rm init}=0.9$~\MS\ or less gives 1/10, \ie one LMCB out of ten
observed should originate from a binary with $M_{1,\rm init}=1.1$~\MS\ and
the rest from the binaries with $M_{1,\rm init}=0.9$~\MS\ or slightly less
if some of them are metal poor. Correcting the average total age of our
models for this factor gives the expected age of about 9~Gyr for the whole
population of LMCBs, which is in a very good agreement with the dynamical
age of 8.9~Gyr obtained by Bilir \etal (2005) for W~UMa type binaries with
periods 0.2--0.3~days. The weighted mean duration time in contact is
0.8~Gyr.

\subsection{Space Density of LMCB and the Frequency of Mergers}
The observations of LMCBs, together with the present models, make possible
an estimate of their space density and the expected rate of merging these
binaries into rejuvenated solar type stars (blue stragglers).

Rucinski (2007) analyzed the space distribution of contact binaries with
different periods, detected within the ASAS program. He concludes that the
density is equal to 14.7 contact binaries with orbital periods between 0.2
and 0.575~days per $10^6$~pc$^3$. Of these, 8.2 binaries have periods
shorter than 0.3~d, which is about 60\% of all W~UMa-type stars. This
translates into 34 LMCB in the immediate solar neighborhood
\ie in the sphere with the radius of 100~pc. The average life time
in contact of LMCB is about 0.8~Gyr (see above). Assuming a stationary
situation, 34 LMCB must appear every 0.8~Gyr in this sphere to replace the
merging binaries, so the average merging rate is 34/0.8~Gyr
$\approx42$~Gyr$^{-1}$. How many blue stragglers resulting from the
coalescence of LMCB is expected in the solar neighborhood? According to our
results LMCB are old, with the average age of 9~Gyr, so only few of them
were formed earlier than, say 8~Gyr ago. Assuming that the presently
estimated rate has existed for the last 2--3~Gyr whereas it was negligible
earlier, we come to about 100 solar type mergers lying within 100~pc from
the Sun. Mergers resulting from binaries with the extreme initial mass
ratio of 3 or more were not included into this order of magnitude estimate
but, unless the formation rate of binaries increases rapidly with the
increasing mass ratio, including them will not change the result
substantially.

Following the coalescence of a binary, an excess AM should be carried away
in the form of the excretion disk of which a part will fall again onto the
star but the rest may stay as a long living Keplerian disk (Zuckerman \etal
2008, Tylenda \etal 2011). Such disks can be the place of planet formation
(Melis \etal 2010, Martin, Spruit and Tata 2011). The planets formed in
disks ejected during the coalescence of a contact binary will be much
younger than the parent stars (Martin, Spruit and Tata (2011).

\subsection{Uncertainties}
It is known from observations of cool contact binaries that numerous
processes and phenomena are present in these systems, which modify their
physical parameters and influence their evolution but, so far, are poorly
understood and difficult to include into the current models. The light
curves of W~UMa-type systems show asymmetric maxima (O'Connell effect)
interpreted as resulting from dynamical phenomena connected with the mass
and energy flow between the components. The shapes of the light curves and
the average brightness level vary from one season to another, sometimes
with an amplitude comparable to the depth of minima (Rucinski and Paczynski
2002, Gazeas, Niarchos and Gradoula 2006, Pilecki 2009). The apparent
surface brightness of the low mass component is higher than that of the
massive component in the majority of W~UMa-type systems (W-phenomenon),
which is contrary to what is expected from simple thermodynamic
considerations of the energy transfer between the components. Orbital
periods of many (if not all) W~UMa-type systems show secular
variations. The rates of these variations are statistically distributed
symmetrically around zero value and have a distribution well described by a
normal curve, indicating their random character (Kubiak, Udalski and
Szymański 2006, Pilecki 2009). Their values are typically very small and
are detectable only thanks to their accumulated random-walk deviations. In
addition, the majority (if not all) of W~UMa-type systems may have
companions causing cyclic period variations through the third-body orbital
perturbations (Pribulla and Rucinski 2006). It seems that the importance of
the period variations for modeling evolution of these stars has often been
overstated.

Most of the above mentioned phenomena are closely related to the magnetic
activity of the components. This activity does not influence conditions
deep in the stellar core, in particular the nuclear energy generation, but
it can substantially modify the outer layers and the surface
conditions. Suddenly appearing dark spots effectively block a part of the
energy flux radiated by a star, which results in an apparent decrease of
the stellar brightness and temperature (a disappearance of spots acts in
the opposite direction). If spots exist longer than the thermal time scale
of the convection zone, the internal structure of this zone is rebuilt. In
equilibrium, the whole core energy flux is radiated away but the
temperature of the inter-spot photosphere increases as also does the
stellar radius (Spruit and Weiss 1986). High level of magnetic activity is
the most likely explanation of a systematic difference between the observed
and modeled stellar radii of late G, K and M type, rapidly rotating
dwarfs. The observed radii show an excess up to 10\% compared to models
(Torres, Anderson and Gimen\'ez 2010). A similar excess is expected in
active binaries. None of the existing evolutionary models of cool contact
binaries includes this effect properly. It should be stressed here that the
rate of evolution in contact depends highly on the accurate values of
stellar radii of both components. An ambiguity of several percent can (and
probably does) influence the secular mass transfer rate in phase~III, hence
its duration (see, for example, Eq.~(13) in Eggleton and Kiseleva-Eggleton
2002). The basic properties of the presented models and their evolution
would not be affected much by ignoring the influence of the magnetic
activity, because they result from processes going on deep in the stellar
interior. However, surface parameters, like stellar radii, apparent
temperatures or the mass transfer rate may depend on the magnetic
phenomena. Winds from both components are expected to interact with one
another, which can also have an effect on the behavior of a contact binary.

Additional uncertainties are connected with the mass and AM loss model used
in this investigation. The uncertainty of the numerical coefficient in the
expression for AML rate, given by Eq.~(1), is about 30\% and that for mass
loss rate appearing in Eq.~(2) is about a factor of two. The AML rate is
saturated in our models for orbital periods shorter than 0.4~d. This is
based on the fact that single stars rotating faster than that show, so
called, supersaturation effect, visible in the X-ray flux (Randich \etal
1996, St\c epie\'n. Schmitt and Voges 2001). Orbital periods of all the contact
models discussed in this paper are significantly shorter than 0.4~d. If the
higher, unsaturated AML rate applies for these binaries, the duration of
phase~III could be shorter by a factor of two.

Nonetheless, the basic ingredients of the present model are quite robust
because they are based on the well investigated process of stellar
evolution and the presence of AML due to magnetized winds. Yet some of the
numerical relations and simplified assumptions of our model need to be
verified with more accurate data. Till then, the quantitative results of
the present paper, in particular the obtained initial parameter range of
the progenitors of LMCB and the calculated duration of the contact
configuration are susceptible to future refinement.

\subsection{Conclusions}
\hglue -7pt This work presents the results of modeling the low mass contact binaries
(LMCB) and their comparison with observations. It is argued that W~UMa-type
systems with orbital periods shorter than about 0.3~d have a few common
properties which are not shared by their counterparts with longer
periods. In particular, they have the total mass lower than 1.4~\MS\ and
the component masses lower than 1~\MS, moderate mass ratios clustering
around 0.5 with lack of very low values close to, or less than 0.1 and
stellar radii placing both components on the MS. They also have low orbital
AM -- less than $3\times10^{51}$ in cgs units. In addition, a great
majority of LMCBs show the W-phenomenon, in contrast to W~UMa-type binaries
with longer periods where a large fraction of variables show
A-phenomenon. LMCBs are intrinsically faint which makes accurate
observations difficult. This is why only few have well determined
parameters (see Table~1), although they make up majority of all W~UMa-type
stars in the solar vicinity (Rucinski 2007). If so, they are the main
source of mergers producing blue stragglers with properties of MS, solar
type stars. After coalescence a circumstellar disk containing the excess AM
of the merger is expected to form. It can be the place for planet formation
(Melis \etal 2010, Martin, Spruit and Tata 2011).

The models of LMCB were obtained from close detached binaries with the
total initial masses less than 1.6~\MS\ and the initial orbital periods
between 1.5 and 2.5~d. Only the proper combination of the initial
parameters results in the formation of LMCB. The condition is that the
massive component must reach its Roche lobe when it is more than halfway
towards TAMS but still some distance to TAMS. If the RLOF occurs when it
has not yet evolved from ZAMS significantly, the rapid mass exchange
results in an immediate merger of both components with no stable contact
phase. If the massive component overflows the Roche lobe when it is close
to TAMS, a new born contact binary has a period longer than 0.3~d and it
later evolves towards the extreme mass ratio. If RLOF occurs when the
massive component has already a small helium core, a short period Algol is
formed following the rapid mass exchange. The comparison of model
calculations with observations shows a very good agreement between the
predicted and observed binary parameters.

The model calculations show that LMCB have the average age of about 9~Gyr
although the binary spends most of its life as detached. Duration of the
contact phase is rather short -- about 0.8~Gyr, \ie of the order of 10\% of
the total life. It follows from the observations and our results that about
30--40 LMCBs and about 100 products of the mergers of these stars are
expected to exist within 100~pc from the Sun.

Our model describes only the evolution of both components and their
orbit. We did not model the energy flow in the contact configuration, so we
can not discuss the component temperatures and, in particular, the
W-phenomenon. This problem was discussed at length by St\c epie\'n (2009). We
only note here that the highest amplitudes of light variations due to the
variable spottiness are observed in single stars with masses 0.7--0.9~\MS,
rotating with periods 0.2--0.3~d (Messina \etal 2003). This is just the
range where the most of the present primaries of LMCB fit. Very high
spottiness of these components may be the reason for the ubiquitous
presence of W-phenomenon which is commonly explained as resulting from a
heavy coverage of the high mass components by dark spots (Hendry, Mochnacki
and Collier Cameron 1992).

The present results are based on a limited number of models with several
simplifying assumptions. Many more models with various initial conditions,
calculated with the modern evolutionary code of interacting binaries are
needed to precisely limit the parameters of the progenitors of LMCB, to
reproduce the observed period distribution, to obtain the accurate
coalescence rate and to fix the precise mass range of the mergers. Such
data have an important significance for the understanding of the planet
formation process, particularly the, so called, hot Jupiters.

On the other hand, the good observational data, particularly spectroscopic,
are still lacking. The picture is muddled by many biases and selection
effects. There are very few systematic surveys, like photometric ASAS or
David Dunlap Observatory radial velocity survey carried out by
S.M.\ Rucinski and collaborators. More such data are urgently needed.

\Acknow{We thank Slavek Rucinski and Wojtek Dziembowski for the very 
careful reading of the paper, comments and corrections.}


\begin{references}
\refitem{Bilir, S., Karata\c{s}, Y., Demircan, O., and Eker, Z.}{2005}{\MNRAS}{357}{497}
\refitem{Binnendijk, L.}{1966}{Publications of DAO}{13}{27}
\refitem{Binnendijk, L.}{1970}{Vistas Astron.}{12}{217}
\refitem{Bradstreet, D.H.}{1985}{\ApJS}{58}{413}
\refitem{Christopoulou, P.-E., Papageorgiou, A., and Chrysopoulos, I.}{2011}{\AJ}{142}{99} 
\refitem{Djura\v{s}ević, G., Yilmaz, M., Ba\c{s}t\"urk, \"O., Kili\c{c}o\u{g}lu, T., Latković, O., and \c{C}ali\c{s}kan, \c{S}.}{2011}{\AA}{525}{66}
\refitem{Eaton, J., Wu, C.C., and Rucinski, S.M.}{1980}{\ApJ}{239}{919}
\refitem{Eggleton, P.P.}{1983}{\ApJ}{268}{368}
\refitem{Eggleton, P.P., and Kiseleva-Eggleton, L.}{2002}{\ApJ}{575}{461}
\refitem{Flannery, B.P.}{1976}{\ApJ}{205}{217}
\refitem{Gazeas, K.D., and Niarchos, P.G.}{2006}{\MNRAS}{370}{L29}
\refitem{Gazeas, K.D., Niarchos, P.G., and Gradoula, G.-P.}{2006}{Astrophys. and Space Sci.}{304}{125}
\refitem{Gazeas, K., and St\c epie\'n, K.}{2008}{\MNRAS}{390}{1577}
\refitem{Girardi, L., Bressan, A., Bertelli, G., and Chiosi, C.}{2000}{\AAS}{141}{371} 
\refitem{Gray, J.D., Woissol, S., and Samec, R.G.}{1996}{IBVS}{~}{4359}
\refitem{Hendry, P.D., Mochnacki, S.W., and Collier Cameron, A.}{1992}{\ApJ}{399}{246}
\refitem{Hrivnak, B.J., Guinan, E.F., and Lu, W.}{1995}{\AJ}{455}{300}
\refitem{Kaluzny, J., and Rucinski, S.}{1986}{\AJ}{92}{666}
\refitem{Khajavi, M., Edalati, M.T., and Jassur, D.M.Z.}{2002}{Astrophys. Space Sci.}{282}{645}
\refitem{Kippenhahn, R., and Weigert, A.}{1967}{Zs. f. Ap.}{65}{251}
\refitem{Kubiak, M., Udalski, A., and Szymański, M.}{2006}{\Acta}{56}{253}
\refitem{Lapasset, E., Gomez, M., and Farinas, R.}{1996}{\PASP}{108}{332}
\refitem{Lee, J.W., Kim, S.-L., Lee, C.-U., and Youn, J.-H.}{2009}{\PASP}{121}{1366} 
\refitem{Lucy, L.B.}{1976}{\ApJ}{205}{208}
\refitem{Maceroni, C., Milano, L., Russo, G., and Sollazzo, C.}{1981}{Astrophys. Space Sci.}{45}{187}
\refitem{Maceroni, C., Milano, L., and Russo, G.}{1984}{\AAS}{58}{405}
\refitem{Maceroni, C., and van't Veer, F.}{1996}{\AA}{311}{523}
\refitem{Marino, B.F., Walker, W.S.G., Bembrick, C., and Budding, E.}{2007}{Publ. of the Astron. Soc. of Australia}{24}{199}
\refitem{Martin, E.L., Spruit, H.C., and Tata, R.}{2011}{\AA}{535}{A50}
\refitem{Melis, C., Gielen, C., Chen, C.H., Rhee, J.H., Song, I., and Zuckerman, B.}{2010}{\ApJ}{724}{470} 
\refitem{Messina, S., Pizzolato, N., Guinan, E.F., and Rodono, M.}{2003}{\AA}{410}{671} 
\refitem{Mochnacki, S.W.}{1981}{\ApJ}{245}{650}
\refitem{Niarchos, P.G., Hoffmann, M., and Duerbeck, H.W.}{1997}{\AAS}{124}{291}
\refitem{Pilecki, B.}{2009}{~}{~}{PhD Thesis, Warsaw University}
\refitem{Pilecki, B., and St\c epie\'n, K.}{2012}{IBVS}{~}{~}{6012}
\refitem{Pojmański, G.}{2002}{\Acta}{52}{397}
\refitem{Pribulla, T., and Rucinski, S.M.}{2006}{\AJ}{131}{2986}
\refitem{Qian, S.-B., Yuan, J.-Z., Soonthornthum, B., Zhu, L.-Y., He, J.-J., and Yang, Y.-G.}{2007}{\AJ}{671}{811}
\refitem{Rahunen, T.}{1981}{\AA}{102}{81}
\refitem{Rahunen, T.}{1982}{\AA}{109}{66}
\refitem{Rahunen, T.}{1983}{\AA}{117}{235}
\refitem{Randich, S., Schmitt, J.H.M.M., Prosser, C.F., and Stauffer, J.R.}{1996}{\AA}{305}{785} 
\refitem{Robb, R.M.}{1992}{IBVS}{~}{~}{3798}
\refitem{Robb, R. M., Greimel, R., and Oullette, J.}{1997}{IBVS}{~}{~}{4504}
\refitem{Rucinski, S.M.}{2000}{\AJ}{120}{319}
\refitem{Rucinski, S.M.}{2006}{\MNRAS}{368}{1319}
\refitem{Rucinski, S.M.}{2007}{\MNRAS}{382}{393}
\refitem{Rucinski, S.M.}{2010}{~}{~}{in ``International Conference on Binaries'', Eds. V. Kalogera and M. van der Sluys, {\it AIP Conf. Proc.}, {\bf 1314}, p.~29}
\refitem{Rucinski, S.M., and Paczynski, B.}{2002}{IBVS}{~}{~}{5321}
\refitem{Rucinski, S.M., and Pribulla T.}{2008}{\MNRAS}{388}{1831}
\refitem{Samec, R.G., Su, W., and Dewitt, J.R.}{1993}{\PASP}{105}{1441}
\refitem{Samec, R.G., Carrigan, B., and Padgen, E.E.}{1995}{IBVS}{~}{~}{4167}
\refitem{Samec, R.G., and Terrell, D.}{1995}{\PASP}{107}{427}
\refitem{Samec, R.G., Gray, J.D., and Carrigan, B.J.}{1995}{\PASP}{107}{136}
\refitem{Samec, R.G., Carrigan, B. J., and Wei Looi, M.}{1998}{\AJ}{115}{1160}
\refitem{Samec, R.G., and Corbin, S.F.}{2001}{IBVS}{~}{~}{5145}
\refitem{Samec, R., and  Loflin, T.}{2004}{Observatory}{124}{284}
\refitem{Samec, R.G., Martin, M., and Faulkner, D.R.}{2004}{~}{~}{5527}
\refitem{Spruit, H.C., and Weiss, A.}{1986}{\AA}{166}{167}
\refitem{St\c epie\'n, K.}{1980}{\Acta}{30}{315}
\refitem{St\c epie\'n, K.}{1995}{\MNRAS}{274}{1019}
\refitem{St\c epie\'n, K.}{2004}{~}{~}{in ``Stars as Suns: Activity, Evolution and Planets'', {\it IAU Symp.}, {\bf 219}, Eds. A.K. Dupree and A.O. Benz, {\it Astr. Soc. of Pacific}, p.~967}
\refitem{St\c epie\'n, K.}{2006}{\Acta}{56}{199}
\refitem{St\c epie\'n, K.}{2009}{\MNRAS}{397}{857}
\refitem{St\c epie\'n, K.}{2011}{\Acta}{61}{139}
\refitem{St\c epie\'n, K., Schmitt, J.H.M.M., and Voges, W.}{2001}{\AA}{370}{157}
\refitem{Szymański, M., Kubiak, M., and Udalski, A.}{2001}{\Acta}{51}{259}
\refitem{Tassoul, J.-L.}{1992}{\ApJ}{389}{375}
\refitem{Torres, G., Andersen, J., and Gim\'enez, A.}{2010}{\AA Rev.}{18}{67}
\refitem{Tylenda, R., Hajduk, M., Kamiński, T., Udalski, A., Soszyński I., Szymański, M.K., Kubiak, M., Pietrzyński, G., Poleski, R., Wyrzykowski, Ł., and Ulaczyk, K.}{2011}{\AA}{528}{114}
\refitem{Vilhu, O.}{1981}{Astrophys. Space Sci.}{78}{401}
\refitem{Vilhu, O.}{1982}{\AA}{109}{17}
\refitem{Webbink, R.F.}{1977}{\ApJ}{211}{881}
\refitem{Webbink, R.F.}{2003}{~}{~}{in `` Stellar Evolution'', Eds. S. Turcotte \etal {\it ASP Conf. Ser.}, {\bf 293}, p.~76}
\refitem{Yakut, K., and Eggleton, P.P.}{2005}{\ApJ}{629}{1055}
\refitem{Yang, Y.-G., Wei, J.-Y, and Li, H.-L.}{2010}{New Astronomy}{15}{155}
\refitem{Zhu, L., Qian, S.-B., Mikul\'a\v{s}ek, Z., Zejda, M., Zv\u{e}\u{r}ina, and Diethelm, R.}{2010}{\AJ}{140}{215}
\refitem{Zola, S., Kreiner, J. M., Zakrzewski, B., Kjurkchieva, D.P., Marchev, D.V., Baran, A., Rucinski, S.M., Ogloza, W. Siwak, M., Koziel, D., Drozdz, M., and Pokrzywka, B.}{2005}{\Acta}{55}{389}
\refitem{Zola, S., Gazeas, K., Kreiner, J.M., Ogloza, W., Siwak, M., Koziel-Wierzbowska, D., and Winiarski, M.}{2010}{\MNRAS}{408}{464}
\refitem{Zuckerman, B., \etal}{2008}{\ApJ}{683}{1085}
\end{references}
\end{document}